\documentclass{aa}
\usepackage{graphicx}
\begin{document}
   \title{ Chemical abundances of Damped Ly $\alpha$ Systems:}
   \subtitle{ A new method for  estimating dust depletion effects }

   \author{Giovanni Vladilo
          \inst{1}
          }

   \offprints{G. Vladilo}

   \institute{Istituto Nazionale di Astrofisica,  
             Osservatorio Astronomico di Trieste,
             Via G.B. Tiepolo 11, I-34131 Trieste\\
              \email{vladilo@ts.astro.it}
             }

   \date{Received ... ; accepted ... }

   \abstract{
A new method is presented for recovering the abundances
of Damped Ly $\alpha$ systems (DLAs) 
taking into account the effects of dust depletion.  
For the first time,
possible variations of  the chemical composition  of the dust
are taken into account in estimating the depletions.  
No prior assumptions  on the  extinction properties of the dust are required. 
%
The method requires a set
of abundances measured in the gas and a set of parameters describing 
the chemical properties of the dust.  
A large subset of these parameters is determined from interstellar observations;
the others are free   parameters for which an educated guess can be made. 
The  method is able to recover the  abundances of the SMC 
 starting from SMC interstellar
measurements apparently discrepant from the stellar ones. 
Application of the method
to 22 DLAs with available [Fe/H] and [Si/Fe] measurements gives the
following results:
(1) the mean  metallicity of the corrected data is 
$< \mathrm{[Fe/H]}> \, \simeq -1.0 $ dex, about 0.5 dex higher than that
of the original data;
(2) the slope of the [Fe/H] versus redshift relation is steeper
for the corrected data ($m \simeq -0.3$ dex) than for the original
ones ($m \simeq -0.2$ dex); 
(3) the corrected [Si/Fe] ratios are less enhanced, on  average, than those
found in Galactic   stars of similar, low metallicity;
(4) a   decrease of the  [Si/Fe] versus [Fe/H] ratios, expected
by "time delay" models of chemical evolution, is found for the corrected data;   
(5) the [Si/Fe] ratios tend to increase with redshift once corrected;
(6) consistency between [Si/Fe] and [S/Zn] measurements,
two independent estimators of the $\alpha$/Fe ratio,
is found only for the corrected abundances.
   }

   \maketitle
%

\section{Introduction}

High resolution spectroscopy of   QSO  absorption line systems
is a very powerful tool for probing structures in the early universe, 
including galaxies in the faint end
of the luminosity function. 
With this technique, elemental abundances can be measured 
in     galaxies up to redshift $z \approx 5$ and used to probe
  the early stages of chemical enrichment. 
In fact,    abundance measurements have been obtained for
a relatively large number of
damped Ly $\alpha$ systems (DLAs), 
the class of QSO absorbers 
most clearly associated with intervening galaxies 
(Lu et al. 1996, 1998; Pettini et al. 1997, 1999, 2000; 
Prochaska \& Wolfe 1999; 
Centuri\'on et al. 1998, 2000; Molaro et al. 2000; Petitjean et al. 2000;
Prochaska et al. 2001).

Abundance measurements can be rather accurate in DLAs:
  column densities of \ion{H}{i} and of  metal species
can easily be determined with errors $\leq 0.1$ dex or even 
$\ll 0.1$ dex;
ionization corrections are generally negligible for most of the
elements considered  
(Vladilo et al. 2001; see also Izotov et al. 2001).
However, the final accuracy of DLA abundances can be severely
affected by the effects of dust. 
Evidence of dust in DLAs comes from extinction   and  abundance studies.
The reddening of QSOs lying behind DLA absorption is
a clear sign  of dust extinction 
(Pei et al. 1991; Pei \& Fall 1995). 
Evidence of elemental depletion   similar to that observed
in the nearby ISM  (Savage \& Sembach 1996) was first reported
by Pettini et al. (1994) from the   behaviour of the
Zn/Cr ratio in DLAs. 
The   enhancement of the Zn/Cr  ratio
relative to the solar value was interpreted as being due to
a different degree of incorporation 
into dust grains of zinc, a volatile element, and chromium,
a refractory element.  
Similar results have been found   for other pairs of
elements with differing degrees of affinity with dust, e.g. Zn/Fe. 
Recent abundance studies  give further support to
the presence of  dust   depletion in DLAs
(Hou et al. 2001, Prochaska \& Wolfe 2002),
indicating that   abundances taken at face value
are of little use 
in understanding the chemical properties of DLA galaxies.

Various approaches have been used to circumvent  the problem of dust
depletion in studies of DLAs abundances. 
One way is to  focus  on elements essentially undepleted,
such as N (Lu et al. 1998, Centuri\'on et al. 1998),
O (Molaro et al. 2000, Dessauges-Zavadsky et al. 2001;
Levshakov et al. 2002),
S (Centuri\'on et al. 2000), 
and Zn (Pettini et al. 1997, 1999; Prochaska \& Wolfe 1999, 2001;
Vladilo 2000). 
Another is to study DLAs with low dust content
(Pettini et al. 2000; Molaro et al. 2000, L\'opez et al. 2002). 
Although these studies do not depend, essentially, on the   
  properties of the dust,  their application 
is limited to certain elements and
to   particular sub-samples of DLAs.  
The only way to perform a general study
of DLAs abundances is to quantify the effects of dust depletion.

By comparing the abundance ratios observed in DLAs  
with those measured in the interstellar gas of the Galaxy
it is possible to derive some indication on the 
depletions of DLAs.
Studies of this kind show a similarity to the depletions
typical of Galactic  warm gas 
(Lauroesch et al. 1996; Kulkarni et al. 1997; 
Savaglio et al. 2000).
In this type of work, two different types of intrinsic DLA abundances  
have been considered: solar abundances and abundances typical of metal-poor stars
in the Milky Way. 
Clearly, it is rather difficult to disentangle  the
intrinsic abundance patterns of DLAs with this type of approach. 

A method for recovering the intrinsic  abundances in individual systems
was presented by Vladilo (1998, hereafter Paper I).  
In that work the chemical composition and extinction properties
of the dust were assumed to be the  
 same  as those  of the dust in Galactic warm gas.
The amount of dust extinction for a given level of metallicity, 
i.e. the dust-to-metals ratio, was  allowed to vary among
 DLAs. 
The dust-to-metals ratio  
of individual DLAs was  estimated
assuming that zinc tracks the iron-peak
elements and that the observed overabundances of Zn/Cr
or Zn/Fe 
are  due entirely to differential depletion.  
A  
limitation of that method is that the dust composition 
may actually vary in different environments,
even within the Milky Way (Savage \& Sembach 1996).
In addition, some recent
investigations  indicate that zinc may not  
exactly follow other iron-peak elements 
(Primas et al. 2000; Umeda \& Nomoto, 2001). 
According to Prochaska \& Wolfe (2002), the assumptions adopted
in Paper I may force the dust-corrected abundances to become
closer to solar values.
To overcome these limitations, a new procedure for estimating
dust depletion in DLAs is presented in this paper.
The equations of the procedure are
based on a new expression for scaling  
dust  depletions    up or down
according to changes of the physical/chemical conditions
of the interstellar gas
 (Vladilo 2002, hereafter Paper II).
This expression allows the dust chemical composition to vary  
according to changes of the dust-to-metals ratio and of
the intrinsic    abundances of the medium. 
With this new formulation,
 deviations of the intrinsic Zn/Fe ratios from solar
values can be consistently accounted for.
At variance with Paper I,
the present method does not make any assumption concerning the extinction
properties of the dust in DLAs. 
The method is presented in Section 2, some
examples of application  are described in Section 3,
and the results are discussed and summarized in Section 4.


\section{The method}

The starting point of the method  is  the expression 
for scaling   depletions up or down derived in Paper II:
\begin{equation}
\label{ScalingLawExtragal}
 f_{\mathrm{X},j}   
= \left( { r_j \over r_i } \right) ^{(1+\eta_\mathrm{X})}
  10^{\left( \varepsilon_\mathrm{X} - 1 \right) 
\left[ { {\rm X} \over {\rm Y} } \right]_j } 
 \, f_{\mathrm{X},i} 
~ ~,
\end{equation}
where  
$f_\mathrm{X} = N_\mathrm{X,dust}/N_\mathrm{X,medium}$  
is the fraction  in dust of the element X; 
the index "medium" refers
to the   totality of the atoms (gas   and  dust);
the index $i$ indicates a  phase of the Galactic ISM 
for which $f_{\mathrm{X},i}$   can  be determined empirically;
 the index $j$ the ISM of an external galaxy
(for instance, a DLA system); 
Y  is an element used   as a reference for
expressing relative abundances. 
The scaling law (\ref{ScalingLawExtragal})
is derived assuming that the abundances in the dust,
$p_\mathrm{X} \equiv (N_\mathrm{X}/N_\mathrm{Y}) _\mathrm{dust}$,
are a function 
of the dust-to-metals ratio, $r$,  and of the intrinsic abundances,
$a_\mathrm{X} \equiv (N_\mathrm{X}/N_\mathrm{Y}) _\mathrm{medium}$, i.e.
\begin{equation}
\label{Assumption}
p_\mathrm{X} = p_\mathrm{X}(r,a_\mathrm{X})
~.
\end{equation}
The dust-to-metals ratio is defined  as the fraction in dust of the
reference element, i.e. $r \equiv f_\mathrm{Y}$.
At variance with Paper I,
the extinction properties of dust grains do not appear in this definition.  
In Eq. (\ref{ScalingLawExtragal})
the intrinsic abundances 
$(N_\mathrm{X}/N_\mathrm{Y}) _\mathrm{medium}  $
are accounted for in the term  
$[ { {\rm X} \over {\rm Y} } ] 
\equiv \log   (N_\mathrm{X}/N_\mathrm{Y}) _\mathrm{medium}  
- \log ( { {\rm X} / {\rm Y} } )_{\sun}$.
The parameters $\eta_\mathrm{X}$
and $\varepsilon_\mathrm{X}$ are
dimensionless derivatives of $p_\mathrm{X}$ 
with respect to  
$r$  and  to $a_\mathrm{X}$, 
respectively.
Given the function (\ref{Assumption}), 
the scaling law (\ref{ScalingLawExtragal})
is always correct for infinitesimal changes of $r$  and  $a_\mathrm{X}$.
For discrete variations of these parameters, the law   is valid if 
$\eta_\mathrm{X}$
and $\varepsilon_\mathrm{X}$ are constant. 
Comparison with the observations indicates that
    a   set of {\em constant} parameters   
is  able to   model all the typical   depletion
patterns   of the Galactic ISM by
only varying $r$  in Eq.  (\ref{ScalingLawExtragal}). 
By allowing both $r$ and $a_\mathrm{X}$ to vary,
  the few extragalactic patterns observed so far\footnote{
Only some interstellar lines of sight to the Magellanic Clouds 
are currently available.}
can be modeled as well. We refer to Paper II for more details. 

The   depletion patterns that can be successfully 
modeled by Eq. (\ref{ScalingLawExtragal})
originate in different types of interstellar gas,
characterized by a variety of  physical conditions
(e.g. cold gas vs. warm gas), metallicities
(e.g. 1 vs. $\simeq$ 0.25 of the solar level),
locations (e.g., Galactic halo vs. disk),
and extinction properties (e.g., strong vs. weak 217.5 nm emission bump).
For these reasons we feel justified in applying the same scaling law
 to the ISM of distant galaxies, including DLAs. 

\subsection{Equations of the method}

The relation between the  observed and the intrinsic abundances 
can be derived from the  Eq.
(\ref{ScalingLawExtragal}) above and from Eq. (4) of Paper I:
\begin{equation}
\label{X_H}
\left[ { \mathrm{X} \over \mathrm{H} } \right]_{\mathrm{obs}} =
\left[ { \mathrm{X} \over \mathrm{H} } \right]_{j} +
\log \left\{
1 -    \varrho ^{(1+\eta_\mathrm{X})}
\, 10^{\left( \varepsilon_\mathrm{X} - 1 \right) 
\left[ { {\rm X} \over {\rm Y} } \right]_{j} } 
\,  f_{\mathrm{X},i} 
\right\}
\,
\end{equation}
where 
$\varrho = r_j/r_i$ 
is the dust-to-metals ratio normalized to $r_i$, 
[X/H]$_{\mathrm{obs}}$   the abundance observed in the gas phase, [X/H]$_{j}$ the
intrinsic abundance and
$[ { {\rm X} \over {\rm H} } ] 
\equiv \log   (N_\mathrm{X}/N_\mathrm{H})  
- \log ( { {\rm X} / {\rm Y} } )_{\sun}$. 

In order to recover the intrinsic abundances we first need to determine $\varrho$.
To do this, we use the expression 
\begin{equation}
\label{DustToMetalsRatio}
\varrho 
-
{  
f_{\mathrm{X},i} \, 
10^{\left[ { \mathrm{X} \over \mathrm{Y} } \right]_j
\varepsilon_\mathrm{X}}
\over 
f_{\mathrm{Y},i} \, 
10^{\left[ { \mathrm{X} \over \mathrm{Y} } \right]_{ \mathrm{obs}} }  
}
\varrho^{ \left( 1+\eta_\mathrm{X} \right) }
+
{
10^{\left[ { \mathrm{X} \over \mathrm{Y} } \right]_j } 
-
10^{\left[ { \mathrm{X} \over \mathrm{Y} } \right]_{\mathrm{obs}} } 
\over
f_{\mathrm{Y},i} \, 
10^{\left[ { \mathrm{X} \over \mathrm{Y} } \right]_{\mathrm{ obs}} }  
}
=
0  ~,
\end{equation}
which can be derived by applying Eq. (\ref{X_H}) 
to the generic element X and to the reference element Y.  
In practice, the solution of  Eq. (\ref{DustToMetalsRatio})
 requires  a measurement of  [X/Y]$_\mathrm{obs}$ and an educated guess of 
[X/Y]$_{j}$.   

Once $\varrho$ is known, we can determine the  relative abundance 
in the medium $[  \mathrm{X} / \mathrm{Y}  ]_\mathrm{j}$
from the expression
\begin{equation}
\label{DustCorrection}
10^{\left[ { \mathrm{X} \over \mathrm{Y} } \right]_\mathrm{j}}
-
\varrho^{\left( 1+\eta_\mathrm{X} \right)} \, f_{\mathrm{X},i} \,
 10^{\left[ { \mathrm{X} \over \mathrm{Y} } \right]_\mathrm{j}{\varepsilon_\mathrm{X}}} 
+  
\left( 
\varrho f_{\mathrm{Y},i} - 1
\right)
 10^{\left[ { \mathrm{X} \over \mathrm{Y} } \right]_\mathrm{obs}}
=0  ~,
\end{equation}
which can be  derived
by rearranging Eq. (\ref{DustToMetalsRatio}). 
Clearly, this second expression can be applied to 
elements X different from the one used  in Eq. (\ref{DustToMetalsRatio}).

Eqs. (\ref{DustToMetalsRatio}) and (\ref{DustCorrection})  
can be solved analytically only in particular cases,
depending on the values 
of the parameters  $\eta_\mathrm{X}$ and $\varepsilon_\mathrm{X}$.
These  equations  have the general form
\begin{equation}
\label{general_equation}
x + a x^\alpha + b = 0 ~,
\end{equation} 
and can be solved by iterations 
given  the coefficients $a$ and $b$ 
and the exponent $\alpha$. In general, the uniqueness of
the solution is not guaranteed. If multiple solutions are present,  
one needs to restrict the search for the roots
within a limited interval of $x$. This interval has to be justified
on physical grounds.   

In the case of Eq. (\ref{DustToMetalsRatio})
 $x = \varrho$ and $\alpha = 1+\eta_\mathrm{X}$.
The exponent $\alpha$
 can attain values higher than unity (see adopted values
of $\eta_\mathrm{X}$ in Table 1). 
A unique solution is found 
 searching for the roots in  the interval  
$0 < \varrho \leq \varrho_\mathrm{lim}$, where $\varrho_\mathrm{lim}=1.0649$ 
is the dust-to-metals ratio corresponding to a fraction of iron in dust
of 100\%. Values outside of this interval are obviously unphysical.   

In the case of Eq. (\ref{DustCorrection})
$x = 10^{\left[ { \mathrm{X} \over \mathrm{Y} } \right]}$ and 
$\alpha=\varepsilon_\mathrm{X}$. 
As explained below, a choice of the exponent in the range
$0 \leq \varepsilon_\mathrm{X} \leq 1$  allows us to probe two extreme
possibilities concerning the dependence of the dust composition   on the
composition of the medium. For these values of   $\alpha$
the solutions of  Eq. (\ref{DustCorrection}) are   unique
and there is no need to restrict the search for the roots
to a particular interval of 
intrinsic abundances $x= 10^{\left[ { \mathrm{X} \over \mathrm{Y} } \right]}$.

 \begin{center} 
\begin{table} 
\caption{ Adopted interstellar parameters }
\begin{tabular}{lll}
\hline \hline
Element & ~~~~~~$f_\mathrm{X}$  & ~~~~~~$\eta_\mathrm{X}$ $^a$ \\  
\hline 
Si & $0.691	\pm 0.069$ & $+4.68 \pm 1.77$ \\      
Mg & $0.715	\pm 0.056$ & $+3.60 \pm 1.50$ \\   
Mn & $0.880	\pm 0.019$ & $+0.53 \pm 0.41$ \\ 
Cr & $0.920	\pm 0.010$ & $+0.37 \pm 0.20$ \\
Fe & $0.939	\pm 0.004$ & $+0.00 \pm 0.07$ \\
Ni	& $0.941	\pm 0.005$ & $+0.01 \pm 0.11$ \\ 
Zn & $0.587	\pm 0.048$ $^{b}$ & $+3.97 \pm 1.56$  $^b$ \\  
\hline
\end{tabular}
\scriptsize{
\\ $^a$ 
The $\eta_\mathrm{X}$ parameters have been determined
by comparing the fractions in dust of the warm gas with those of the cold
gas in the Galactic disk.
\\
$^b$ 
Warm disk fraction in dust not available for zinc; the adopted 
values are based on the observed zinc depletions
in the cold disk and in the warm disk+halo gas scaled up or down
with relation (\ref{ScalingLawExtragal}).}
\end{table}
\end{center}

\subsection{Input parameters}

To apply the procedure various sets of interstellar parameters are
required.   
The  fractions in dust  $f_{X,i}$ and
the parameters $\eta_\mathrm{X}$ can be determined from 
Galactic interstellar studies, as shown in Paper II. 
In Table 1 we list the set of  $f_{X}$ and  $\eta_\mathrm{X}$ parameters
that allow one to model simultaneously all types of Galactic depletion
patterns by only varying $\varrho=r_j/r_i$. 
The Galactic depletion patterns include  the {\em cold disk},
{\em warm disk}, {\em warm disk+halo}, and {\em warm halo} ones, as defined by
Savage \& Sembach (1996)\footnote{
The   depletions
from Savage \& Sembach (1996) have been updated according to revised values
of \ion{Mg}{ii} (Welty et al. 1999 and refs. therein)
and \ion{Ni}{ii} (Fedchak \& Lawler 1999; Howk et al. 1999)
oscillator strengths. Zinc has been added to the original list
by including   measurements   of
 the same   lines of sight selected by
 Savage \& Sembach  (Roth \& Blades 1995). 
Solar reference values from Anders \& Grevesse (1989) are adopted
consistently   in the present paper.  
}.

In principle, the $\varepsilon_\mathrm{X}$ parameters can be determined  
from extragalactic interstellar observations.  
The preliminary study of the Small Magellanic Cloud 
described in Paper II 
indicates that
the observed depletions can be reproduced for 
$\varepsilon_\mathrm{X} \simeq 1$, or somewhat below unity. 
However,  this  result  could be specific to the few
extragalactic lines of sight investigated. 
Therefore, in the present work 
we prefer  to consider $\varepsilon_\mathrm{X}$ as a free
parameter.
Luckily, an educated guess of   $\varepsilon_\mathrm{X}$  can be made
considering the physical meaning of this parameter.  
As shown in Paper II,  ${\delta p_\mathrm{X} \over p_\mathrm{X} }
\simeq \varepsilon_\mathrm{X} { \delta a_\mathrm{X} \over a_\mathrm{X} } $.
In other words, $\varepsilon_\mathrm{X}$ indicates 
the sensitivity of the abundance ratio X/Y in the dust
to changes of the same ratio in the medium.   
The case $\varepsilon_\mathrm{X}=1$ is what one may expect on 
intuitive grounds: the more abundant an element is,
the more atoms of that element will be
available for being incorporated into the dust.    
However, given the complexity of the processes involving dust formation,
accretion and destruction,  
the composition of the medium might play a secondary role in
determining the composition of the dust, at least for some
elements, in which case $\varepsilon_\mathrm{X} \approx 0$. 
Values $\varepsilon_\mathrm{X} > 1$ do not have a special appeal
from the intuitive point of view and are not supported by the
interstellar study of the SMC carried out in Paper II.  
Based on these considerations, we   adopt  here  
$\varepsilon_\mathrm{X}=1$ or   $\varepsilon_\mathrm{X}=0$ 
as two limiting possibilities.
The diversity (similarity) between the results obtained in these two cases    
indicates the weakness (robustness) of the method due to the poor knowledge of  
the function $p_\mathrm{X} = p_\mathrm{X}(a_\mathrm{X})$. 
Given the form of the equations 
(\ref{X_H}), (\ref{DustToMetalsRatio}) 
and (\ref{DustCorrection}),
the uncertainty of
$\varepsilon_\mathrm{X}$ will significantly affect  the results only if the intrinsic
abundances   deviate significantly  from solar ratios.

\begin{flushleft} 
\begin{table} 
\caption{ Sets of  [Zn/Fe]$_j$ and $\varepsilon_\mathrm{X}$ parameters }
\begin{tabular}{cccc}
\hline \hline
Set  & [Zn/Fe]$_j$  & $\varepsilon_\mathrm{Zn}$ & $\varepsilon_\mathrm{X}$ \\
\hline
S00 & +0.0 & 0 & 0 \\
S10 & +0.0 & 1 & 0 \\
S01 & +0.0 & 0 & 1 \\
S11 & +0.0 & 1 & 1 \\
E00 & +0.1 & 0 & 0 \\
E10 & +0.1 & 1 & 0 \\
E01 & +0.1 & 0 & 1 \\
E11 & +0.1 & 1 & 1 \\
\hline
\end{tabular}
\end{table}
\end{flushleft}

\subsubsection{ The intrinsic [Zn/Fe] ratio }

Interstellar studies indicate that the fraction of zinc in dust  
  is relatively small while iron is  
heavily depleted (Savage \& Sembach 1996).  
On the other hand,
studies of Galactic  stars indicate that 
the zinc abundance  nearly  tracks  
that  of iron-peak elements
down to metallicities [Fe/H] $\simeq -2.2$ dex 
(Sneden et al. 1991).
Should   zinc     track   iron in DLAs 
  precisely
(i.e.,  [Zn/Fe]$_{j}$ =0),
the [Zn/Fe] ratio  would be a perfect indicator of depletion
and   an optimal choice for determining the
dust-to-metals ratio with Eq. (\ref{DustToMetalsRatio}). 
However,  Galactic stars 
show a significant enhancement of the ratio  [Zn/Fe] 
at metallicities
[Fe/H] $< -2.2$ dex (Primas et al. 2000). 
This result suggests
a difference in the nucleosynthetic origin of zinc and iron
(Umeda \& Nomoto 2002)
and makes  the use of the [Zn/Fe] ratio
as an indicator of depletion less reliable
(Prochaska \& Wolfe 2002). 
The metallicities of DLAs are generally higher than $-2.2$ dex.
In this range the  [Zn/Fe] ratio in Galactic stars
is consistent with the solar value within the errors,
with a scatter of about $\pm 0.1$ dex
at 1$\sigma$ level and   tentative evidence for an enhancement
of $\simeq + 0.1$ dex.  
Therefore, we may expect that the  enhancement of
the intrinsic ratio [Zn/Fe]$_{j}$ in DLAs is little or negligible
in the same range of metallicities.  
The DLAs with the lowest metallicity and dust content should
reveal an intrinsic enhancement, if present.
The best example at our disposal, 
the  DLA at $z=3.39$ towards Q0000-26, with 
[Zn/H] $\simeq -2.1$ dex
(Molaro et al. 2000),
 shows very little evidence for [Zn/Fe] enhancement. 
In the rest of this paper
we consider   [Zn/Fe]$_{j}$ as an input parameter
to which we assign values 0.0 or +0.1 dex
in the solution of our problem.

\subsection{Abundances of individual velocity components}
 
The absorption spectrum of DLAs is often characterized
by the presence of multiple components.
These may be associated 
with parcels of gas located in different
regions of the intervening galaxy. 
The method presented here can be applied to   the
analysis of  individual components.
However, while the velocity components of   metal lines
may be   
resolved in high resolution spectra,  those of   \ion{H}{i} lines 
are not. 
This implies that Eqs. (\ref{DustToMetalsRatio}) and (\ref{DustCorrection})
 can be applied to individual components,
but Eqs.  (\ref{X_H}) cannot. 
In other words, relative abundances and dust-to-metals ratios can
be obtained for individual components. The absolute abundances 
can be corrected for dust effects only as far as the integrated
absorption is concerned (i.e. the total column densities inside the DLA).
This contraint is set by the nature of the
measurements   and not by a theoretical limitation 
 of the method.

\section{Applications of the method}

Some examples of application of the method  are presented
in this section. 
First, the method is tested in a nearby  galaxy, namely the Small Magellanic
Cloud.  The SMC is the only external galaxy
with several accurate interstellar measurements available and
with  intrinsic abundances    known from stellar observations. 
We show that the method is successful in recovering the intrinsic SMC
abundances, a result that gives us more confidence in applying the 
same procedure
in high redshift DLAs with unknown   stellar abundances. 
In     Section 3.2 
the method is   used to correct   [Fe/H] metallicities
and [Si/Fe] ratios in DLAs.

In all the applications presented here Fe is  adopted
as the reference element Y. 
The Zn/Fe ratio is  used to determine the 
normalized dust-to-metals ratio $\varrho$ 
with Eq. (\ref{DustToMetalsRatio}). 
Errors in $\varrho$ are estimated by taking into account the  
measurement errors
of the \ion{Zn}{ii} and \ion{Fe}{ii} column densities.
The dust-corrected abundances are obtained by solving 
Eq. (\ref{DustCorrection}) analytically for 
$\varepsilon_\mathrm{X} = 0$ and $=1$.  
The errors of the dust-corrected abundances 
shown in Table 3 and in the figures  
are derived from the propagation of the $\varrho$ errors
and of  the  
[X/Fe]$_\mathrm{obs}$ or [Fe/H]$_\mathrm{obs}$ errors. 
The adopted  parameters   $f_\mathrm{X}$ and $\eta_\mathrm{X}$
derived from interstellar studies
are listed in Table 1. 
The    sets of educated guesses of the 
[Zn/Fe]$_j$ and $\varepsilon_\mathrm{X}$ parameters are shown in Table 2. 
Since  we are   interested in
deriving the {\em absolute} abundance of the reference element,  [Fe/H],
the analysis of individual velocity components is not considered 
(see Section 2.5).

\subsection{ SMC interstellar lines of sight }

\begin{figure}
  \centering
  \includegraphics[width=8.5cm]{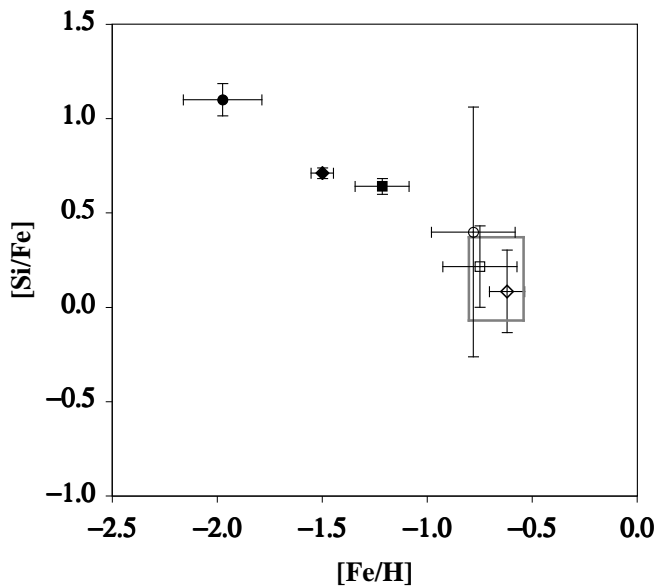}
 \caption{ Interstellar [Si/Fe] versus [Fe/H] in three
lines of sight of the SMC
(squares: Sk 78; diamonds: Sk 108; circles:  Sk 155). 
Filled symbols: observed values taken from Welty et al. (2001).
Open symbols: values  corrected for dust effects using the
set of parameters E11. 
Box: range of [Si/Fe] vs. [Fe/H] values obtained from SMC
stellar observations (Russell \& Dopita 1992).  
  }
         \label{fSiFe}
 \end{figure}

The interstellar lines of sight of the SMC considered here are those
towards the stars Sk 78, Sk 108, and Sk 155, for
which accurate abundance measurements are available (see Paper II).  
In Fig. \ref{fSiFe} we plot   the [Si/Fe] versus [Fe/H]
results for these three SMC lines of sight. 
For the sake of clarity, only one set of Table 2 input parameters 
is considered in the figure.
The uncorrected abundances (filled symbols) suggest the existence of
 a spread of the [Fe/H] metallicity  and of the [Si/Fe] ratio
among the different lines of sight. However,
none of them is in agreement with the abundances
representative of the  SMC 
(box in the figure). 
The corrected ratios (open symbols) yield instead consistent results, all 
in good agreement with the intrinsic  SMC ratios. 
Since the three different lines  of sight are treated independently,
each   with its own value of the $\varrho$ parameter,
it is remarkable that they
give consistent
results only when the dust correction is applied.

Similar results are obtained from the study of the
[Mg/Fe], [S/Fe], [Cr/Fe], [Mn/Fe], and [Ni/Fe]
ratios in the same SMC lines of sight. 
These different   ratios   are corrected
independently of each other, since
each element X has its own value of $\eta_\mathrm{X}$.
Therefore, it is remarkable that each one of them  
matches the corresponding SMC [X/Fe] ratio only  when the dust
correction is applied.  
The results are not significantly affected by the choice of the input
parameters listed in Table 2.  

In summary, the interstellar abundances not corrected for
dust depletion do not match the intrinsic SMC abundances.
The corrected abundances yield consistent results for
different elements and different lines of sight, all in
agreement with the intrinsic SMC abundances.
Most of the results are modestly affected by the choice of
the    [Zn/Fe]$_\mathrm{j}$   and  
$\varepsilon_\mathrm{X}$ parameters.

\subsection{ Dust-corrected  abundances in DLAs }

Iron is often used as a tracer of metallicity  
in DLAs because many \ion{Fe}{ii} transitions 
with different oscillator strengths can be detected 
in a broad range of absorption redshift.  
Silicon is the only $\alpha$-capture element which has been extensively 
observed in DLAs, with over 30 measurements available.
Therefore,
the Si/Fe ratio is the only 
abundance ratio of $\alpha$-capture over iron-peak elements 
for which  a  statistically significant 
sample can be analyised. 
As a first application of the new method to DLAs we present here
a list of dust-corrected  [Fe/H] and [Si/Fe] abundances. 
Only DLAs with good quality measurements of \ion{H}{i},
\ion{Si}{ii}, \ion{Fe}{ii}, and \ion{Zn}{ii} have been considered.
The Zn data are required to derive the dust-to-metals ratio $\varrho$. 
The resulting sample of 22 systems is shown in Table 3 together with 
the relevant references for the adopted abundances. 
The [Fe/H] and [Si/Fe] measurements with and without dust correction 
are compared in the table.  
Four sets of corrected abundances are shown, corresponding to
the sets of input parameters  
  S00, S11, E00, and E11  in Table 2.
The results  are not 
strongly affected by the choice of these parameters. 
The values of $\varrho$ obtained   
from the [Zn/Fe] data
indicate that the level of depletion is always smaller than that found
in the cold gas of the Galactic disk. 
In fact, the median $\varrho$ value of the sample is 0.74 and 0.61
for [Zn/Fe]$_j = 0.0$ and $+0.1$ dex, respectively. 
These values of $\varrho$ imply that 
the  iron fraction in dust   in DLAs
is   lower, on average, than 
that  in Galactic warm halo gas, $\varrho_\mathrm{wh}=0.84$ (Paper II). 
The broad similarity between depletion in DLAs and depletions
in warm halo gas found in previous work
(Lauroesch et al. 1996; Savaglio et al. 2000)
is therefore confirmed by the present analysis.  
In the rest of this section we present some general conclusions derived
from the dust-corrected sample of Table 3. 

\subsubsection{ Metallicity }

As in the case of SMC interstellar data, also in DLAs 
the use of [Fe/H] without dust correction can lead
to a severe underestimate of the real metallicity of the system.
As shown in Table 3, the difference between
between original and corrected measurements can be as high as 
1 dex in some cases. 
The statistical properties of the sample are severely affected
by this effect.
For instance, the mean metallicity of the original measurements  is 
$<\mathrm{[Fe/H]}> \simeq -1.5$ dex, but  that of the corrected data is
$<\mathrm{[Fe/H]}> \simeq -1.0$ dex. 
Also  metallicity versus redshift studies   are affected. 
For instance, a linear regression of [Fe/H] versus $z_\mathrm{abs}$  
for the sample of Table 3 yields
 slopes\footnote{
The slope is obtained from the least-squares method and is defined as 
$m = { \sum_i (z_{\mathrm{abs},i} - <z_\mathrm{abs}>) 
(\mathrm{[Fe/H]}_i - <\mathrm{[Fe/H]}>)
\over \sum_i (z_{\mathrm{abs},i} - <z_\mathrm{abs}>)^2 }$,
where $<z_\mathrm{abs}>$ and $<\mathrm{[Fe/H]}>$ are the mean values of
the sample.
}
$m \simeq -0.32$ and $\simeq -0.17$ for the data with
and without dust correction, respectively.
The slope is steeper for the corrected data,
as expected if dust depletion is more important at higher
metallicities.
The slope $m \simeq -0.3$ of the corrected metallicities is similar to that  
found  
by Savaglio et al. (2000) and by Vladilo et al. (2000).
In the former work  a large  sample of  absorbers
was corrected for dust effects
assuming that the intrinsic abundance ratios in DLAs are solar. 
In the work by Vladilo et al. (2000) the uncorrected [Zn/H] ratio
was used as a dust-free indicator of metallicity. 
While the interpretation of any possible trend between
metallicity and redshift is very complex, it is
clear that  studies of this type based on [Fe/H]
data not corrected for dust  should be considered with  caution. 

  \begin{figure}
  \centering
  \includegraphics[width=8.5cm]{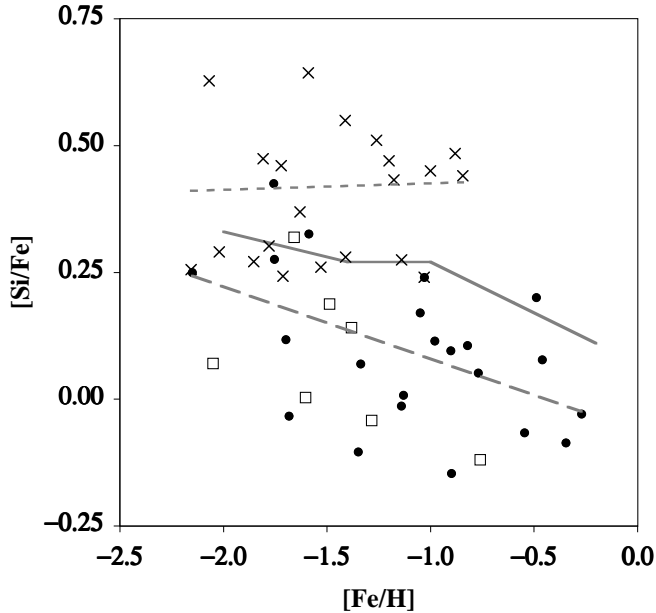}
 \caption{  
[Si/Fe] and [Fe/H]  ratios in DLAs. 
Crosses: original  measurements. 
Dotted line: linear regression through the original measurements.  
Filled circles: measurements corrected with the set of
parameters E11 (Table 2).
Dashed line: linear regression through the dust-corrected data. 
Open squares: [S/Zn] versus [Zn/H] data in DLAs. 
Continuous line: midmean vector of [Si/Fe] versus [Fe/H]   in Galactic
metal-poor stars (Ryan et al. 1996). 
  }
         \label{fstat}
 \end{figure}

  \begin{figure}
  \centering
  \includegraphics[width=8.5cm]{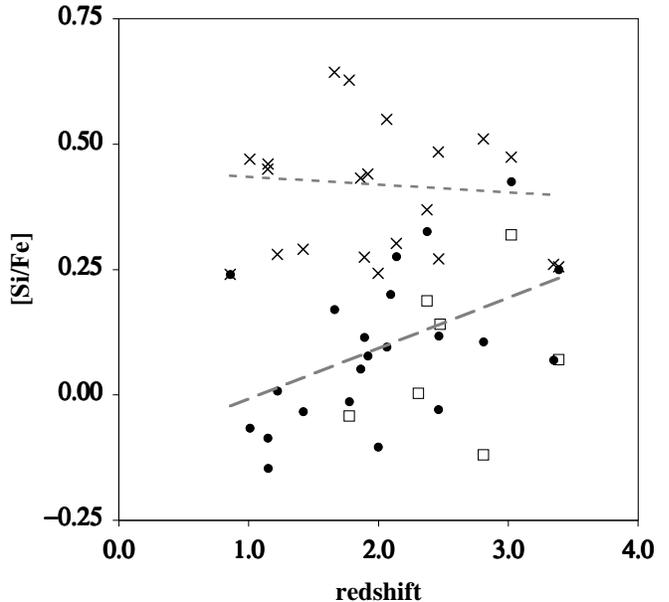}
 \caption{  
 [Si/Fe] ratios versus absorption redshift  in DLAs. 
Crosses: original  measurements. 
Dotted line: linear regression through the original measurements.  
Filled circles: measurements  corrected with the set of
parameters E11 (Table 2).
Dashed line: linear regression through the dust-corrected
 data. 
Open squares: [S/Zn] ratios versus absorption redshift in DLAs.
  }
         \label{SiFe&redshift}
 \end{figure}

\subsubsection{ [Si/Fe] abundance ratios}

The abundance ratio of $\alpha$-capture over iron-peak elements
is an   indicator of  chemical evolution. 
In Galactic stars of low metallicity
the $\alpha$/Fe ratio is  enhanced relative to the solar value
and  decreases with increasing [Fe/H] 
(e.g. Ryan et al. 1996). This behaviour can be 
naturally explained by   models of
 galactic evolution   in terms of  a "time delay" between
the enrichment of $\alpha$ elements  
and that of Fe-peak elements  
(e.g. Matteucci 1991 and refs. therein).
 
Observed trends of $\alpha$/Fe  ratios in DLAs, particularly the Si/Fe ratio,
have been used to characterize the evolutionary status of DLA galaxies
(Lu et al. 1996; Prochaska \& Wolfe 1999, 2002;
Pettini et al. 2000; Molaro et al. 2000).
The [Si/Fe] ratios taken at face value are significantly enhanced at low metallicity
($\simeq +0.4$ dex), and this fact  was originally
taken as an indication that DLAs may be
undergoing a chemical evolution similar
to that of a young Milky Way (Lu et al. 1996). 
Subsequent work on dust free $\alpha$/Fe ratios
(Centuri\'on et al. 2000; Molaro et al. 2000)
and on dust-corrected Si/Fe ratios (Paper I)
 has seriously questioned this point of view.
The enlargement of the abundance data base
gives now a convincing evidence that the [Si/Fe] enhancement is
largely due to dust depletion, as shown for instance by
the behaviour of the [Si/Fe] versus [Zn/Fe] ratios 
(Prochaska \& Wolfe 2002).
In any case,  as we show in Fig. \ref{fstat}, both the original
[Si/Fe] measurements and the   corrected ones indicate a 
behaviour of DLAs different
from that typical of metal-poor Galactic stars. In fact,
most of the non-corrected data lie {\em above} the "midmean vector" representative
of Galactic stars (continuous line; Ryan et al. 1996),  
while most of the corrected data lie {\em below}.
This is true for all the different sets of input parameters
considered in Table 2.
The filled circles  in Fig. \ref{fstat} indicate the data corrected
with the set of parameters E11. 
Very similar results are obtained for the other sets of parameters  
with enhanced [Zn/Fe]$_j$ ratio (E00, E01, E10). 
The corrected data lie even more below the Galactic stars adopting the sets of
parameters with solar [Zn/Fe]$_j$ ratio (S00, S01, S10, S11).

The amount of [Si/Fe] enhancement at the lowest 
DLA metallicities
($< -1.5$ dex) is a probe of
the {\em earliest stages} of the chemical evolution.  
 From the analysis of the [Si/Fe] versus [Si/H] data,
Prochaska \& Wolfe (2002) find evidence for a [Si/Fe] plateau 
([Si/Fe] $\simeq +0.3$ dex)
at [Si/H] $< -1.5$ dex, 
that they attribute to the lack of dust depletion at low
metallicities. 
The present analysis  indicates that 
2 out of 6 DLAs with metallicities  $< -1.5$ dex have
non-negligible [Si/Fe] depletion (Table 3). 
This indicates that some systems may be depleted 
even in the Prochaska \& Wolfe's [Si/Fe] plateau. 

In addition to the degree of enhancement, the possible presence
of a slope in the [Si/Fe] data gives   important information on
the chemical evolution.  
Independent of the specific prescriptions of the models, 
the time delay between the injection of $\alpha$ elements
and that of Fe-peak elements implies that the
$\alpha$/Fe ratio must decrease with increasing [Fe/H]. 
The [Si/Fe] measurements in DLAs taken at face value do not show such a
trend. The untilted dotted line in Fig. \ref{fstat}
represents the linear regression through the non-corrected data. 
On the other hand,  the   corrected ratios do show
evidence for a decrease of   [Si/Fe] with [Fe/H].
As an example, the tilted dashed line in  Fig. \ref{fstat} represents
the linear regression through the data corrected
with the E11  parameters. 
The slope  resembles that of metal-poor stars,
even if the ratios are lower on the average. 

Once corrected for dust depletion, the [Si/Fe] ratios
show a tendency to increase with redshift
which is not seen in the original measurements
(Fig. \ref{SiFe&redshift}). 
The typical value is [Si/Fe] $\approx 0.0$ dex
at $z \simeq 1$ and increases up to
[Si/Fe] $\approx +0.25$ dex at $z \simeq 3.5$, 
with a large dispersion at all redshifts.   
This possible trend of decreasing [Si/Fe] with cosmic time  is   
consistent with the general expectations of the time delay models.
Since DLAs probably represent an heterogeneous collection of
different types of galaxies, 
the interpretation of this result is quite complex  
 and will be considered in a separate work. 
In any case,
the fact that only the dust-corrected ratios show
the expected trends    is rather encouraging.  

An important support to the validity of the correction procedure
comes from the comparison with [S/Zn] data.
Since S and Zn are almost unaffected by depletion,
the S/Zn ratio can be considered a sort of dust free 
$\alpha$/Fe ratio (Centuri\'on et al. 2000).  
If the method presented here works properly,  we expect
that the   corrected [Si/Fe] ratios and the non-corrected [S/Zn] ratios
show a similar behaviour.\footnote
{
Apart  from little differences 
that may be present if S does not track perfectly Si,
and Zn does not follow exactly Fe.
}
The limited sample of [S/Zn] data (Lu et al. 1996; Kulkarni et al. 1996;
Molaro et al. 1998; Prochaska \& Wolfe 1999; Centuri\'on et al. 2000)
 shown as open squares in
Figs.  \ref{fstat} and \ref{SiFe&redshift}, confirms this expectation.
The non-corrected [Si/Fe] ratios are generally inconsistent
with the [S/Zn] data.  

\scriptsize{
\begin{center} 
\begin{table*} 
\scriptsize{
\caption{ [Fe/H] and [Si/Fe] abundance ratios in DLAs: observed and dust-corrected
values$^a$  }
\begin{tabular}{lcrrrlrrrr}
\hline \hline
Identifier	&	$z_\mathrm{abs}$	&	
[Zn/H]$_\mathrm{obs}$			&	[Fe/H]$_\mathrm{obs}$			&	[Si/Fe]$_\mathrm{obs}$			&	Refs$^b$
& 	 
 [Fe/H]$_\mathrm{cor}^{c}$			&	 [Fe/H]$_\mathrm{cor}^{c}$		
&	 [Si/Fe]$_\mathrm{cor}^{c}$				&	  [Si/Fe]$_\mathrm{cor}^{c}$			\\
\hline
	&		&				&				&				&		 		&	S00 \& S11			&	E00 \& E11			&	 S00 \& S11			& E00 \& E11			\\
\hline
0000-263	&	3.3901	&	-2.05	$\pm$	0.09	&	-2.16	$\pm$	0.09	&	0.26	$\pm$	0.04	&	Mol\&00	&	-2.05	$\pm$	0.10	&	-2.15	$\pm$	0.09	&	0.15	$\pm$	0.07	&	0.25	$\pm$	0.05	\\
	&		&				&				&				&		&				&	-2.15	$\pm$	0.09	&	0.15	$\pm$	0.07	&	0.25	$\pm$	0.05	\\
0149+33	&	2.1400	&	-1.65	$\pm$	0.14	&	-1.78	$\pm$	0.10	&	0.30	$\pm$	0.05	&	P\&W99	&	-1.65	$\pm$	0.15	&	-1.75	$\pm$	0.12	&	0.18	$\pm$	0.12	&	0.28	$\pm$	0.08	\\
	&		&				&				&				&		&				&	-1.75	$\pm$	0.12	&	0.18	$\pm$	0.12	&	0.28	$\pm$	0.08	\\
0201+365	&	2.4620	&	-0.27	$\pm$	0.06	&	-0.88	$\pm$	0.04	&	0.48	$\pm$	0.01	&	P\&W96	&	-0.09	$\pm$	0.10	&	-0.30	$\pm$	0.08	&	-0.07	$\pm$	0.01	&	-0.01	$\pm$	0.03	\\
	&		&				&				&				&		&				&	-0.27	$\pm$	0.09	&	-0.11	$\pm$	0.21	&	-0.03	$\pm$	0.05	\\
0302-223	&	1.0095	&	-0.56	$\pm$	0.12	&	-1.20	$\pm$	0.12	&	0.47	$\pm$	0.06	&	Pet\&00	&	-0.36	$\pm$	0.17	&	-0.58	$\pm$	0.15	&	-0.08	$\pm$	0.03	&	-0.04	$\pm$	0.06	\\
	&		&				&				&				&		&				&	-0.55	$\pm$	0.16	&	-0.14	$\pm$	0.35	&	-0.07	$\pm$	0.12	\\
0347-383	&	3.0250	&	-1.66	$\pm$	0.11	&	-1.81	$\pm$	0.12	&	0.47	$\pm$	0.07	&	Pro\&01	&	-1.66	$\pm$	0.15	&	-1.76	$\pm$	0.14	&	0.32	$\pm$	0.11	&	0.42	$\pm$	0.10	\\
	&		&				&				&				&		&				&	-1.76	$\pm$	0.14	&	0.32	$\pm$	0.11	&	0.42	$\pm$	0.10	\\
0454+039	&	0.8597	&	-1.01	$\pm$	0.11	&	-1.03	$\pm$	0.07	&	0.24	$\pm$	0.10	&	Pet\&00	&	-1.01	$\pm$	0.09	&	-1.03	$\pm$	0.07	&	0.22	$\pm$	0.11	&	0.24	$\pm$	0.10	\\
	&		&				&				&				&		&				&	-1.03	$\pm$	0.07	&	0.22	$\pm$	0.11	&	0.24	$\pm$	0.10	\\
0515-4414	&	1.1510	&	-0.99	$\pm$	0.11	&	-1.72	$\pm$	0.22	&	0.46	$\pm$	0.27	&	dlV\&00	&	-0.70	$\pm$	0.44	&	-0.97	$\pm$	0.37	&	-0.08	$\pm$	0.09	&	-0.08	$\pm$	0.17	\\
	&		&				&				&				&		&				&	-0.90	$\pm$	0.44	&	-0.20	$\pm$	2.52	&	-0.15	$\pm$	1.08	\\
0528-2505	&	2.8110	&	-0.76	$\pm$	0.12	&	-1.26	$\pm$	0.15	&	0.51	$\pm$	0.12	&	Lu\&96	&	-0.67	$\pm$	0.26	&	-0.83	$\pm$	0.22	&	0.01	$\pm$	0.14	&	0.11	$\pm$	0.17	\\
	&		&				&				&				&		&				&	-0.82	$\pm$	0.23	&	0.01	$\pm$	0.15	&	0.11	$\pm$	0.13	\\
0841+129	&	2.3745	&	-1.49	$\pm$	0.10	&	-1.63	$\pm$	0.17	&	0.37	$\pm$	0.15	&	P\&W99	&	-1.49	$\pm$	0.23	&	-1.59	$\pm$	0.20	&	0.23	$\pm$	0.22	&	0.33	$\pm$	0.18	\\
	&		&				&				&				&		&				&	-1.59	$\pm$	0.20	&	0.23	$\pm$	0.22	&	0.33	$\pm$	0.18	\\
1104-1805	&	1.6616	&	-1.02	$\pm$	0.01	&	-1.59	$\pm$	0.02	&	0.64	$\pm$	0.03	&	Lop\&99	&	-0.88	$\pm$	0.05	&	-1.07	$\pm$	0.04	&	0.06	$\pm$	0.03	&	0.16	$\pm$	0.03	\\
	&		&				&				&				&		&				&	-1.05	$\pm$	0.04	&	0.08	$\pm$	0.06	&	0.17	$\pm$	0.03	\\
1117-1329	&	3.3511	&	-1.24	$\pm$	0.10	&	-1.53	$\pm$	0.09	&	0.26	$\pm$	0.06	&	Per\&02	&
-1.23	$\pm$	0.13	&	-1.34	$\pm$	0.12	&	-0.03	$\pm$	0.10	&	0.07	$\pm$	0.10	\\
	&		&				&				&				&		&				&	-1.34	$\pm$	0.12	&	-0.03	$\pm$	0.10	&	0.07	$\pm$	0.10	\\
1210+1731	&	1.8918	&	-0.88	$\pm$	0.10	&	-1.14	$\pm$	0.12	&	0.27	$\pm$	0.07	&	Pro\&01	&	-0.87	$\pm$	0.14	&	-0.98	$\pm$	0.14	&	0.01	$\pm$	0.10	&	0.11	$\pm$	0.10	\\
	&		&				&				&				&		&				&	-0.98	$\pm$	0.14	&	0.01	$\pm$	0.09	&	0.11	$\pm$	0.10	\\
1215+33	&	1.9990	&	-1.27	$\pm$	0.08	&	-1.71	$\pm$	0.09	&	0.24	$\pm$	0.06	&	P\&W99	&	-1.21	$\pm$	0.14	&	-1.35	$\pm$	0.12	&	-0.18	$\pm$	0.06	&	-0.10	$\pm$	0.08	\\
	&		&				&				&				&		&				&	-1.35	$\pm$	0.12	&	-0.20	$\pm$	0.06	&	-0.10	$\pm$	0.08	\\
1223+178	&	2.4661	&	-1.60	$\pm$	0.10	&	-1.85	$\pm$	0.10	&	0.27	$\pm$	0.02	&	Pro\&01	&	-1.59	$\pm$	0.11	&	-1.70	$\pm$	0.11	&	0.02	$\pm$	0.04	&	0.12	$\pm$	0.04	\\
	&		&				&				&				&		&				&	-1.70	$\pm$	0.11	&	0.02	$\pm$	0.04	&	0.12	$\pm$	0.04	\\
1247+267	&	1.2232	&	-1.04	$\pm$	0.26	&	-1.41	$\pm$	0.13	&	0.28	$\pm$	0.16	&	Pet\&99	&	-1.01	$\pm$	0.37	&	-1.13	$\pm$	0.32	&	-0.09	$\pm$	0.27	&	0.01	$\pm$	0.31	\\
	&		&				&				&				&		&				&	-1.13	$\pm$	0.33	&	-0.09	$\pm$	0.23	&	0.01	$\pm$	0.30	\\
1331+170	&	1.7764	&	-1.28	$\pm$	0.05	&	-2.07	$\pm$	0.04	&	0.63	$\pm$	0.00	&	P\&W99	&	-0.94	$\pm$	0.07	&	-1.24	$\pm$	0.06	&	-0.02	$\pm$	0.00	&	0.01	$\pm$	0.01	\\
	&		&				&				&				&		&				&	-1.14	$\pm$	0.07	&	-0.06	$\pm$	0.45	&	-0.01	$\pm$	0.24	\\
1351+318	&	1.1491	&	-0.36	$\pm$	0.16	&	-1.00	$\pm$	0.13	&	0.45	$\pm$	0.16	&	Pet\&99	&	-0.16	$\pm$	0.33	&	-0.38	$\pm$	0.26	&	-0.09	$\pm$	0.08	&	-0.05	$\pm$	0.15	\\
	&		&				&				&				&		&				&	-0.35	$\pm$	0.30	&	-0.16	$\pm$	0.91	&	-0.09	$\pm$	0.30	\\
1354+258	&	1.4200	&	-1.60	$\pm$	0.14	&	-2.02	$\pm$	0.11	&	0.29	$\pm$	0.16	&	Pet\&99	&	-1.55	$\pm$	0.25	&	-1.69	$\pm$	0.21	&	-0.12	$\pm$	0.18	&	-0.03	$\pm$	0.21	\\
	&		&				&				&				&		&				&	-1.68	$\pm$	0.22	&	-0.13	$\pm$	0.17	&	-0.03	$\pm$	0.21	\\
2206-199A	&	1.9200	&	-0.39	$\pm$	0.10	&	-0.84	$\pm$	0.10	&	0.44	$\pm$	0.02	&	P\&W97	&	-0.32	$\pm$	0.11	&	-0.47	$\pm$	0.10	&	-0.02	$\pm$	0.02	&	0.08	$\pm$	0.03	\\
	&		&				&				&				&		&				&	-0.46	$\pm$	0.11	&	-0.02	$\pm$	0.02	&	0.08	$\pm$	0.02	\\
2230+025	&	1.8642	&	-0.70	$\pm$	0.09	&	-1.18	$\pm$	0.09	&	0.43	$\pm$	0.02	&	P\&W99	&	-0.63	$\pm$	0.10	&	-0.78	$\pm$	0.09	&	-0.04	$\pm$	0.03	&	0.05	$\pm$	0.04	\\
	&		&				&				&				&		&				&	-0.77	$\pm$	0.10	&	-0.05	$\pm$	0.02	&	0.05	$\pm$	0.03	\\
2231-0015	&	2.0662	&	-0.86	$\pm$	0.10	&	-1.41	$\pm$	0.12	&	0.55	$\pm$	0.07	&	P\&W99	&	-0.74	$\pm$	0.17	&	-0.92	$\pm$	0.15	&	0.00	$\pm$	0.07	&	0.10	$\pm$	0.09	\\
	&		&				&				&				&		&				&	-0.90	$\pm$	0.16	&	0.00	$\pm$	0.15	&	0.10	$\pm$	0.07	\\
2359-0216	&	2.0950	&	-0.75	$\pm$	0.10	&	-1.67	$\pm$	0.10	&	0.90	$\pm$	0.03	&	P\&W99	&	-0.31	$\pm$	0.12	&	-0.64	$\pm$	0.12	&	0.05	$\pm$	0.01	&	0.12	$\pm$	0.03	\\
	&		&				&				&				&		&				&	-0.49	$\pm$	0.13	&	0.17	$\pm$	1.00	&	0.20	$\pm$	0.68	\\ 
\hline
\end{tabular}
\scriptsize{  
\\ 
$^a$ When necessary, the column densities of the literature have been revised
using the oscillator strengths given by Welty et al. (1999).  
\\
$^b$ References for  the \ion{H}{i}, \ion{Si}{ii}, \ion{Fe}{ii}, and \ion{Zn}{ii}
column densities, with the following exceptions: \\
\ion{H}{i} in Q0000-26 and Q2231-0015 (Lu et al. 1996);  
\ion{Fe}{ii} in Q0841+129 (Centuri\'on et al. 2000); 
\ion{Zn}{ii} in Q0347-383   (Levshakov et al. 2002). 
\\ 
$^c$ For each system, the values in the first row are relative to the sets of
parameters S00 or E00; the values in the second row  to the sets S11 or E11;
the corrected [Fe/H] ratios are equal in the cases S00 and S11; see Table 2.
\\  
\\
References \\ 
dlV\&00: de la Varga et al. 2000;  
L\'op\&99: L\'opez et al. 1999; 
Lu\&96: Lu et al. 1996; 
Mol\&00: Molaro et al. 2000;   
P\'er\&02: P\'eroux et al. 2002; 
Pet\&97: Pettini et al. 1997; 
Pet\&99: Pettini et al. 1999;
Pet\&00: Pettini et al. 2000;  
P\&W96: Prochaska \& Wolfe 1996;
P\&W97: Prochaska \& Wolfe 1997; P\&W99: Prochaska \& Wolfe 1999; 
P\&W00: Prochaska \& Wolfe 2000; Pro\&01: Prochaska et al. 2001.
}
}
\end{table*}
\end{center}

\normalsize

\section{Discussion and summary}

Chemical abundances of DLAs  probe
the effects of nucleosynthesis in high \ion{H}{i} density regions of the
universe seen  at different cosmic epochs throughout
 a considerable fraction of the Hubble time.
However,  
disentangling the effects of nucleosynthesis from those
of dust depletion   is a rather difficult task. 
In this work a new method has been presented for
quantifying dust   effects, with the aim of recovering
the intrinsic abundances starting from the observed ones.
There are several important differences with respect
to the original method presented in Paper I and  previous work on the same subject. 

The first difference concerns the definition
of the dust-to-metals ratio. 
In Paper I this was defined in terms of extinction per unit
level of metallicity, with the aim of estimating {\em indirectly} the    
extinction of DLAs from the elemental depletions measured in
each system. 
The broad agreement  of such indirect estimates
with the direct,  but rather uncertain, estimates  
obtained from QSO reddening studies (Pei et al. 1991)
is a positive test  of consistency of the results obtained
 in Paper I.  
However, in order to compare these two types of estimates  
it is necessary to assume that the  
extinction is proportional to the  amount of dust present in the system. 
This assumption is risky, since the extinction is  also
determined by the geometry and size distribution of the dust grains.   
The present work  focuses only on the {\em abundances} in the dust
and in the medium (gas plus dust).
In order to avoid any assumption on the {\em extinction}  
of the dust, we have defined the dust-to-metals ratio 
as the fraction of atoms  in dust  form of an element Y
chosen as a reference,  i.e. $r=N_\mathrm{Y,dust}/N_\mathrm{Y,medium}$.  
 
The second difference is the way that            
elemental depletions are scaled up or down 
as a function of the dust-to-metals ratio.
In the work by Kulkarni et al. (1997) and in Paper I  
the dust-to-metals ratio was allowed to change, but the    
abundances in the dust were kept constant. 
In this case, the fractions in dust of all elements are scaled up or down
together. Therefore, all    fractions in dust 
are proportional to that of the reference element,
i.e. $f_\mathrm{X} \propto r$.  
In the present work the fraction in dust
of various elements have different dependence on $r$,  
according to the general 
law  $f_\mathrm{X} \propto r^{1+\eta_\mathrm{X}}$.
This law  has been derived and calibrated in Paper II
and is able to reproduce all the  types of depletion patterns
observed in   the Milky Way interstellar gas  
(Savage \& Sembach 1996)
with a single set of $\eta_\mathrm{X}$ parameters.

The third difference is that, for the first time, elemental
depletions are allowed to vary also as a function of the
overall abundances of the medium.  
As shown in Paper II, the dependence of the fraction in dust
on the intrinsic abundances must be of the type
$f_\mathrm{X} \propto  
10^{(\varepsilon_\mathrm{X}-1)  \left[ { \mathrm{X} \over \mathrm{Y} }\right]}$.
The introduction of the parameter   $\varepsilon_\mathrm{X}$ 
allows us to consider different behaviours of the dust composition.
When $\varepsilon_\mathrm{X}=0$, the composition of the dust does not
vary with the composition of the medium. 
This extreme case could be valid, for instance, if the number
ratio of  two elements in the dust  
is {\em uniquely} determined by the 
capability of these two elements
to have chemical bonds.  
In general, however, we may expect that the abundance ratio X/Y
in the dust will track the abundance ratio X/Y in the medium,
in which case $\varepsilon_\mathrm{X}=1$. 
By considering both cases $\varepsilon_\mathrm{X}=0$
and $\varepsilon_\mathrm{X}=1$ we can take into account
 very different behaviours of the dust composition with respect 
to changes of the overall abundances,
without entering into the details of the  complex 
 physical processes  of dust
formation, accretion and destruction. 

As in Paper I, the present method can be applied to individual 
DLA systems. If the quality of the observational data is good
enough, the method can be used for studying the abundances
of distinct velocity components inside a given DLAs.
The main steps of the procedure for recovering intrinsic abundances
can be summarized as follows. 
First, the dust-to-metals ratio  
  is  derived with  Eq. (\ref{DustToMetalsRatio}).
The most natural choice for the [X/Y] ratio to be used
in that equation is the [Zn/Fe] ratio. 
In practice, the dust-to-metals ratio  
is derived  from
the observed  [Zn/Fe] value
and an educated guess of the intrinsic  [Zn/Fe] value.
Since the equations of the new
method  take into account the dependence of the abundances in the dust
on the abundances in the medium, 
possible changes of the intrinsic [Zn/Fe] ratio in DLAs
can now be  treated  self-consistently.
Therefore the unknown value of the intrinsic
[Zn/Fe] ratio  is now considered as an input parameter which, at variance
with Paper I, may also assume values different from zero.  
At this point  the abundances of elements other than Zn
can be recovered from the observed  ones
by means of Eqs. (\ref{X_H}) and (\ref{DustCorrection}).
The dust-corrected abundances can be obtained
for different sets of input parameters.  In this way, we can test
the stability of the results for different values of 
the $\varepsilon_\mathrm{X}$ parameter and of the intrinsic
[Zn/Fe] ratio.  

The    new method has been applied
to correct the few existing   interstellar abundances
measured with accuracy in a nearby galaxy, namely the SMC. 
The results of this exercise  support the validity of the  procedure. 
In fact, for
 all the abundance ratios considered
(Fe/H, Mg/Fe, Si/Fe, S/Fe, Cr/Fe, Mn/Fe, and Ni/Fe)
the  corrected interstellar abundances are in agreement  
with the intrinsic SMC abundances
(known   from stellar data).
The method yields consistent results for  the 
different lines of sight considered. 
The importance of the dust correction procedure
is illustrated by the fact that
the non-corrected interstellar data would  indicate
an SMC metallicity $-2.0 \leq \mathrm{[Fe/H]} \leq -1.2$,
a strong enhancement of the [$\alpha$/Fe]  ratios,
and  a high degree of inhomogeneity of
the SMC. All these indications are contradicted
by stellar data. 

The successful results obtained for a low-metallicity  galaxy
such as the SMC are encouraging since the final goal of
the method is   recovering the abundances of DLA systems,
also characterized by a low metallicity level. 
There are other reasons why
we feel justified in applying the dust correction method
to  DLAs systems. In fact,
 the adopted scaling law   
  (Eq. \ref{ScalingLawExtragal})
can successfully model depletion patterns originated in
a variety of interstellar environments, with different
   physical conditions
(e.g. cold vs. warm), locations (e.g. Galactic halo vs. disk; SMC),
and 
extinction properties (e.g., strong vs. weak 217.5 nm emission bump).

The results obtained by applying the present method to
the sample of DLAs with available \ion{H}{i}, \ion{Si}{ii}, \ion{Fe}{ii},
 and \ion{Zn}{ii} measurements
can be summarized as follows.

The mean metallicity of the sample is
$<\mathrm{[Fe/H]}> \simeq -1.0$ dex for the dust-corrected data, 
significantly  higher than that of 
 the original data, $<\mathrm{[Fe/H]}> \simeq -1.5$ dex.
The difference 
between observed and corrected metallicities can be as high as 
1 dex in some cases. 
Also the [Fe/H] versus redshift relation in DLAs is 
affected by dust effects, the slope being  
$\simeq -0.32$ and $\simeq -0.17$ for the data with
and without dust correction, respectively. 

Also the Si/Fe ratio, the most frequently measured $\alpha$/Fe ratio in DLAs,
is significantly affected by depletion. 
Most of the dust-corrected [Si/Fe] ratios  
lie  below the median values representative of Galactic stars.
The [Si/Fe] enhancement  at low metallicities  ([Fe/H] $\simeq -1.5/-2.0$ dex)
is typically [Si/Fe] $\simeq +0.2$ dex. This value is somewhat lower
than the [Si/Fe]$\simeq +0.3$ dex plateau of the original [Si/Fe] data 
 found by Prochaska \& Wolfe (2002) at the same metallicities, 
 suggesting that  some depletion  is   present 
also in that plateau. 

Possible trends   of   [Si/Fe]   versus [Fe/H] 
and versus redshift are also affected by   depletion effects.  
The   corrected [Si/Fe] data  decrease with increasing [Fe/H]
and   increase with $z_\mathrm{abs}$. Both trends are
qualitatively consistent with the expectations of the time delay
between the injection of $\alpha$-capture elements and that  of Fe-peak
elements, predicted by models of chemical evolution. 
The uncorrected [Si/Fe] ratios  do not show such  trends. 

An important  test  of the dust correction method
is provided by  the comparison between   Si/Fe   and   
  S/Zn data. 
In fact, the S/Zn ratio is an  
indicator of the  $\alpha$/Fe ratio  which does not require dust correction
to be applied since both S and Zn are essentially undepleted in
the interstellar gas.  
The   [Si/Fe] ratios  show a   behaviour similar to that of the
  [S/Zn] ratios only if the corrections are applied.
This result supports the validity of the 
method. 

A detailed interpretation of the 
intrinsic DLA abundances   
is quite complex since DLAs   represent
a heterogeneous, and probably biased, collection of interstellar regions
associated with different types of galaxies.  
A detailed comparison of the   dust-corrected abundances with 
the  predictions of chemical evolution models of different
types of galaxies will be presented in a separate paper (Calura et al. 2002).

\begin{acknowledgements}
     
 I wish to thank Irina Agafonova  for providing routines for
the solution of Eq. (\ref{general_equation}). 
The data for the DLA system
towards BR 1117-1329 have been kindly provided by C\'eline P\'eroux
and collaborators in advance of publication.

\end{acknowledgements}

\clearpage

\end{document}